\documentstyle[psfig,epsf]{article}
\input tcilatex

\begin{document}
\hfill CERN-TH/99-328

\begin{center}
{\Large \bf Exact results on quantum field theories interpolating between
pairs of conformal field theories} \\

\vspace{4mm}

%                      author/address
Damiano Anselmi\\
CERN, Theory Division, CH-1211, Geneva 23\\
Switzerland \\
\end{center}

%                       Abstract
\begin{abstract}
I review
recent results on conformal field theories 
in four dimensions and quantum field theories interpolating between
conformal fixed points, supersymmetric and non-supersymmetric.
The talk is structured in   three parts: {\it i)} central charges, {\it ii)}
anomalous dimensions and  {\it iii) }quantum irreversibility. 
XIV International Workshop on High Energy Physics and Quantum Field Theory,
Moscow, June, 1999.

\end{abstract}
\vspace{4mm}

The family of theories, in general UV-free, that interpolate between two
conformal fixed points, in such a way 
that the IR limit is reachable by (resummed)
perturbation theory, is called {\it conformal window}. The conformal window, which can be viewed as the {\it convergence radius} 
of the perturbative series,
does not contain QCD, where other non-perturbative effects have to be
taken into account. Yet it is the region separating 
the perturbative regime from QCD.
Understanding the conformal window better can be a source of insight into the
low-energy limit of QCD itself.

In two dimensions conformal field theories have an infinite symmetry \cite
{zamolotutti} and are sometimes exactly solvable. In higher dimensions, there
are simplifications in the presence of supersymmetry and exact results are
available. Very general theorems, implications of unitarity, give exact
results even in the absence of supersymmetry.

Here I summarize the research that I undertook on these issues over the
past three years. The paper is divided in three sections: {\it i)} central
charges, based on ref. \cite{noi}; {\it ii)} anomalous dimensions, on refs. 
\cite{n=4,n=2,OPE}; and {\it iii) }quantum irreversibility, on refs. 
\cite{athm,6d,c=a}.

\vskip .3truecm \noindent {\bf 1. Central charges} \vskip .2truecm

\noindent I consider, as a concrete example, N=1 supersymmetric QCD with
group $G=SU(N_{c})$ and $N_{f}$ quarks in the fundamental representation. I
compute the infrared values of the gravitational central charges called $c$
and $a$ in the conformal window $3N_{c}/2<N_{f}<3N_{c}$.

The theory contains gauge superfields $V^{a}$, $a=1,\ldots ,N_{c}^{2}-1$,
and chiral quark and antiquark superfields, $Q^{\alpha i}$ and $\widetilde{Q%
}_{\alpha i}$, $\alpha =1,\ldots ,N_{c}$, $i=1,\ldots ,N_{f}$, whose
physical components are the gauge potentials $A_{\mu }^{a}$ and Majorana
gauginos $\lambda ^{a}$, and the complex scalars $\phi ^{\alpha i}$ and $%
\tilde{\phi}_{\alpha i}$ and Majorana spinors $\psi ^{\alpha i}$ and $\tilde{%
\psi}_{\alpha i}$, respectively. This theory has the usual gauge
interactions and no superpotential.

The Konishi and $R$ currents, whose fermion contributions are 
\begin{equation}
K_{\mu }=\frac{1}{2}\bar{\psi}\gamma _{\mu }\gamma _{5}\psi +\frac{1}{2}%
\widetilde{\bar{\psi}}\gamma _{\mu }\gamma _{5}\widetilde{\psi },\qquad
R_{\mu }=\frac{1}{2}\bar{\lambda}^{a}\gamma _{\mu }\gamma _{5}\lambda ^{a}-%
\frac{1}{6}(\bar{\psi}\gamma _{\mu }\gamma _{5}\psi +\widetilde{\bar{\psi}}%
\gamma _{\mu }\gamma _{5}\widetilde{\psi }),  \nonumber
\end{equation}
are classically conserved, but anomalous at the quantum level. We
distinguish {\it internal} and {\it external} anomalies, the latter
associated with external background sources.

The internal anomalies of $K_{\mu }$ and $R_{\mu }$ are expressed by the
operator equations 
\[
\partial _{\mu }R^{\mu }={\frac{1}{48\pi ^{2}}}[3N_{c}-N_{f}(1-\gamma
)]F_{\mu \nu }^{a}\tilde{F}_{\mu \nu }^{a},\qquad \partial _{\mu }K^{\mu }=%
\frac{N_{f}}{16\pi ^{2}}F_{\mu \nu }^{a}\tilde{F}_{\mu \nu }^{a}. 
\]
There is an anomaly-free, RG-invariant combination of $K_{\mu }$ and $%
R_{\mu }$ \cite{kogan}: 
\[
S_{\mu }=R_{\mu }+\frac{1}{3}\left( 1-\frac{3N_{c}}{N_{f}}-\gamma \right)
K_{\mu }. 
\]
The coefficient of $K_{\mu }$ is the numerator of the exact NSVZ \cite{nsvz} 
$\beta $-function 
\begin{equation}
\beta (g)=-\frac{g^{3}}{16\pi ^{2}}\frac{3N_{c}-N_{f}(1-\gamma (g))}{%
1-g^{2}N_{c}/8\pi ^{2}}  \label{eq:2.5}
\end{equation}
and $\gamma /2$ is the anomalous dimension of the superfield $Q$ (or $%
\widetilde{Q}$).

The first example of exact IR result is the anomalous dimension 
\begin{equation}
\gamma _{{\rm IR}}=1-\frac{3N_{c}}{N_{f}}  \label{eq:2.29}
\end{equation}
of the quark fields, obtained by setting $\beta $ equal to zero. 
We now compute other interesting quantities in
the IR limit.

$R_{\mu }$ is the lowest component of the supercurrent superfield $J_{\alpha 
\dot{\alpha}}$ that also contains the stress tensor and supersymmetry
currents. To study the gravitational central charges we introduce the
background metric $g_{\mu \nu }$ and source $V_{\mu }$ for the $R$-current.
In these background fields the trace and $R$-anomalies are related by
supersymmetry and read, in a critical theory, 
\[
{\Theta }=\frac{c}{16\pi ^{2}}(W_{\mu \nu \rho \sigma })^{2}-\frac{a}{16\pi
^{2}}(\widetilde{R}_{\mu \nu \rho \sigma })^{2}+\frac{c}{6\pi ^{2}}V_{\mu
\nu }^{2},\quad \partial _{\mu }R^{\mu }=\frac{c-a}{24\pi ^{2}}R_{\mu \nu
\rho \sigma }\widetilde{R}^{\mu \nu \rho \sigma }+\frac{5a-3c}{9\pi ^{2}}%
V_{\mu \nu }\widetilde{V}^{\mu \nu }. 
\]
We include a factor of $\sqrt{g}$ in the definition of $R^{\mu }$. Here $%
W_{\mu \nu \rho \sigma }$ is the Weyl tensor and $\widetilde{R}_{\mu \nu
\rho \sigma }$ is the dual of the curvature tensor, the second term of $%
\Theta $ being the Euler density; $V_{\mu \nu }$ is the field strength of $%
V_{\mu }$. The coefficient $a$ of the Euler density is an independent
constant, while the coefficients of the $(W_{\mu \nu \rho \sigma })^{2}$ and 
$(V_{\mu \nu })^{2}$ terms are related. This can be proved by observing that
the two-point function of $J^{\alpha \dot{\alpha}}$ has a unique structure
in superspace \cite{noiprimo}, 
\begin{equation}
\langle J_{\alpha \dot{\alpha}}(z)J_{\beta \dot{\beta}}(0)
\rangle \propto {c}\,\,{\frac{%
s_{\alpha \dot{\beta}}\bar{s}_{\beta \dot{\alpha}}}{(s^{2}\bar{s}^{2})^{2}}},
\label{eq:4.2}
\end{equation}
and calculating the partial derivative $\mu ~\partial /\partial \mu $ of the
correlators 
\[
\langle T_{\mu \nu }(x)\,T_{\rho \sigma }(0)\rangle =-{\frac{c}{48\pi ^{4}}%
\prod }_{\mu \nu ,\rho \sigma }^{(2)}\left( {\frac{1}{|x|^{4}}}\right)
,\qquad \langle R_{\mu }(x)\,R_{\nu }(0)\rangle =\frac{c}{3\pi ^{4}}\ {\pi }%
_{\mu \nu }\left( {\frac{1}{|x|^{4}}}\right) , 
\]
using 
\begin{equation}
\int \Theta =\mu {\frac{\partial }{\partial \mu }},~~~~~~\mu {\frac{\partial 
}{\partial \mu }}\left( {\frac{1}{|x|^{4}}}\right) =2\pi ^{2}\delta (x).
\label{buffo}
\end{equation}
Here ${\pi }_{\mu \nu }=\partial _{\mu }\partial _{\nu }-\Box \delta _{\mu
\nu }$ and ${\prod }_{\mu \nu ,\rho \sigma }^{(2)}=2\pi _{\mu \nu }\pi
_{\rho \sigma }-3(\pi _{\mu \rho }\pi _{\nu \sigma }+\pi _{\mu \sigma }\pi
_{\nu \rho }).$ The second relation of (\ref{buffo}) can be obtained by
means of a regularization technique.

In a free supersymmetric gauge theory with $N_{v}$ gauge and $N_{\chi }$
chiral multiplets, the values of $c$ and $a$ are 
\[
c_{UV}=\frac{1}{24}\left( 3N_{v}+N_{\chi }\right) {},\quad a_{UV}=\frac{1}{48%
}\left( 9N_{v}+N_{\chi }\right) . 
\]
Off-criticality, there are additional terms in $\Theta $ and $\partial _{\mu
}R^{\mu }$, proportional to $\beta (g)$, including the internal
contribution $\beta /4~F_{\mu \nu }^{2}$,
and the central charges depend on the coupling, i.e. $c=c(g)$ and $%
a=a(g)$.

Since $S_{\mu }$ is quantum-conserved in the absence of sources, its
external anomalies are $\mu$-independent \cite{thooft}: 
\[
\partial _{\mu }S^{\mu }=\frac{s_{1}}{24\pi ^{2}}R_{\mu \nu \rho \sigma }%
\widetilde{R}^{\mu \nu \rho \sigma }+\frac{s_{2}}{9\pi ^{2}}V_{\mu \nu }%
\widetilde{V}^{\mu \nu }. 
\]
A one-loop computation gives 
\[
s_{1}=\frac{1}{16}(N_{c}^{2}+1),\qquad s_{2}=\frac{9}{16}\left(
N_{c}^{2}-1-2N_{c}N_{f}\left( \frac{N_{c}}{N_{f}}\right) ^{3}\right) . 
\]
Now, we observe that $R_{\mu }=S_{\mu }$ in the IR limit, whence 
\[
s_{1}=c_{{\rm IR}}-a_{{\rm IR}},~~~~~~s_{2}=5a_{{\rm IR}}-3c_{{\rm IR}}, 
\]
so that we finally get 
\begin{equation}
c_{{\rm IR}}={\frac{1}{16}}\left( 7\,N_{c}^{2}-2-9{\frac{N_{c}^{4}}{N_{f}^{2}%
}}\right) ,\quad \quad a_{{\rm IR}}={\frac{3}{16}}\left( 2\,N_{c}^{2}-1-3{%
\frac{N_{c}^{4}}{N_{f}^{2}}}\right) .  \label{gru}
\end{equation}
Observe that $c_{{\rm IR}}$ and $a_{{\rm IR}}$ are non-negative throughout
the conformal window, in agreement with their nature of central charges. In
particular, the inequality $c_{{\rm IR}}\geq 0$ follows from reflection
positivity of the stress-tensor two-point function.

The total flows of the central charges are 
\begin{equation}
c_{{\rm UV}}-c_{{\rm IR}} =-\frac{N_{c}N_{f}}{48}\gamma _{IR}\left( 3\frac{%
N_{c}}{N_{f}}+9\frac{N_{c}^{2}}{N_{f}^{2}}-4\right) ,~~ 
~~a_{{\rm UV}}-a_{{\rm IR}} =\frac{N_{c}N_{f}}{48}\gamma _{IR}^{2}\left(
2+3\frac{N_{c}}{N_{f}}\right) \geq 0.  \label{fora}
\end{equation}
The difference $a_{{\rm UV}}-a_{{\rm IR}}$ is everywhere positive in the
conformal window, as conjectured by Cardy \cite{cardy}. This phenomenon is
called {\it quantum irreversibility}. Instead, the difference $c_{{\rm UV}%
}-c_{{\rm IR}}$ is positive in part of the conformal window and negative in
the rest.

A corollary of the above derivation is that both $c$ and $a$ are constant on
families of conformal field theories, i.e. they are marginally uncorrected.

The procedure that I have illustrated can be applied any time there is a
unique $R$-current and a conformal window. The effects of mass perturbations
and symmetry breaking can be straightforwardly included. The analysis of a
wide class of models, done in \cite{noi2}, confirms the conclusions just
derived, in particular the inequalities $a_{{\rm UV}}\geq a_{{\rm IR}}\geq 0$%
, on which I have more to say in section 3.

\vskip .3truecm \noindent {\bf 2. Anomalous dimensions} \vskip .2truecm

\noindent The second class of quantities that characterize a conformal field
theory are the anomalous dimensions. Given that the operator-product
expansion of the stress-tensor does not close, of primary interest is the
spectrum of anomalous dimensions of the (infinitely many) higher-spin
currents generated by the singular terms of the
$TT$ OPE. In
general the classification is not simple, but with the help of supersymmetry
we can reach this goal in various models. Using theorems discovered in the
context of the deep inelastic scattering, in particular the
Ferrara--Gatto--Grillo theorem \cite{FGG} and the Nachtmann theorem \cite
{nacht}, several conclusions about strongly coupled conformal field theories
can be derived. These conclusions hold also for non-supersymmetric theories.

The algorithm to work out the currents of the quantum conformal algebra
starts from the stress tensor $T$ and the spin-0 component $\Sigma _{0}$ of
the Konishi multiplet, which is the first operator generated by the $TT$ OPE 
\cite{noiprimo}, and proceeds via a combination of two steps:

$i)$ supersymmetry, which moves ``vertically'' in the algebra, i.e.
changes the dimension of the operators and therefore their
singularity in the OPE;

$ii)$ orthogonalization of two-point functions, which moves ``horizontally'',
i.e. at the same singularity level in the OPE.

I illustrate this very briefly in the case of the N=2 vector multiplet. The
current multiplets have length 2 in spin units, in particular the
multiplet of the stress tensor. There is one multiplet for each spin, even
or odd.

The vector, spinor and scalar contributions to the currents of the N=2
vector multiplet $(A_{\mu }$,$\lambda _{i}$, $M,N)$, $i=1,2$, are
schematically given in the free-field limit by 
\begin{eqnarray*}
{\cal J}^{V} &=&F_{\mu \alpha }^{+}\overleftrightarrow{\Omega }_{{\rm even}%
}F_{\alpha \nu }^{-},\quad {\cal J}^{F}=\frac{1}{2}\bar{\lambda}_{i}\gamma
_{\mu }\overleftrightarrow{\Omega }_{{\rm odd}}\lambda _{i},~~~{\cal J}^{S}=M%
\overleftrightarrow{\Omega }_{{\rm even}}M+N\overleftrightarrow{\Omega }_{%
{\rm even}}N, \\
{\cal A}^{V} &=&F_{\mu \alpha }^{+}\overleftrightarrow{\Omega }_{{\rm odd}%
}F_{\alpha \nu }^{-},\quad {\cal A}^{F}=\frac{1}{2}\bar{\lambda}_{i}\gamma
_{5}\gamma _{\mu }\overleftrightarrow{\Omega }_{{\rm even}}\lambda
_{i},\quad {\cal A}^{S}=-2iM\overleftrightarrow{\Omega }_{{\rm odd}}N,
\end{eqnarray*}
plus improvement terms \cite{high}, where $\overleftrightarrow{\Omega }_{%
{\rm even/odd}}$ denotes an even/odd string of derivative operators $%
\overleftrightarrow{\partial }$, ${\cal J}$,${\cal A}$ denote the even and
odd (axial) currents, and $V,F,S$ mean vector, fermion, scalar. A simple set
of rules determines the operation $(i)$. The result is 
\begin{eqnarray*}
{\cal J}^{S} &\rightarrow &-2{\cal A}^{F}+2{\cal A}^{S},\quad {\cal J}%
^{F}\rightarrow -8{\cal A}^{V}+{\cal A}^{S},\quad {\cal J}^{V}\rightarrow -2%
{\cal A}^{V}+\frac{1}{4}{\cal A}^{F}, \\
{\cal A}^{F} &\rightarrow &-8{\cal J}^{V}+~{\cal J}^{S},\quad {\cal A}%
^{S}\rightarrow -2{\cal J}^{F}+2{\cal J}^{S},\quad {\cal A}^{V}\rightarrow -2%
{\cal J}^{V}+\frac{1}{4}~{\cal J}^{F}.
\end{eqnarray*}
This operation raises the spin by one unit and it is independent of the spin
on the basis $({\cal J}^{S,F,V},$ ${\cal A}^{S,F,V})$, which is, however,
not diagonal in the sense of point $(ii)$. The diagonalization produces the
correct higher-spin currents, which are rational combinations of $({\cal %
J}^{S,F,V},$ ${\cal A}^{S,F,V}).$

The current multiplet of the stress tensor reads in particular 
\[
{\cal T}_{0}={\frac{1}{2}}{\cal J}_{0}^{S},\qquad {\cal T}_{1}=-{\cal A}%
_{1}^{F}+{\cal A}_{1}^{S},\quad \quad {\cal T}_{2}=8{\cal J}_{2}^{V}-2{\cal J%
}_{2}^{F}+{\cal J}_{2}^{S}.
\]
It contains also a spin-1 current ${\cal T}_{1}$ (an $R$-current) and a
spin-0 mass operator ${\cal T}_{0}$.

One can proceed similarly for the hypermultiplet and then combine the two in
an N=2 finite theory, which is the case we are interested in here. This
family is parametrized by a coupling constant $g$ as well as the rank $N_{c}$
of the gauge group, which we assume to be $SU(N_{c})$. Multiplets having
different minimal spins are orthogonal, but some pairs of multiplets
have the same minimal spin. These, in general, mix under renormalization.
In particular, there is a multiplet ${\cal T}^{*}$ mixing with ${\cal T}$.

At $g$ different from zero the higher-spin currents acquire anomalous
dimensions (and are extended to include other supersymmetric partners that
disappear when $g=0)$. Let $h_{2s}$ denote the minimal anomalous dimensions
of the even-spin levels. The Ferrara--Gatto--Grillo--Nachtmann (FG\-GN) theorem
states that, starting with the spin-2 level, the spectrum $h_{2s}$
is positive, increasing and convex:

\[
0\leq h_{2s}\leq h_{2(s+1)},~~~~~h_{2(s+1)}-h_{2s}\leq h_{2s}-h_{2(s-1)}. 
\]
The most important implication of this theorem is that the OPE algebra
generated by the multiplet of the stress tensor {\it does} close, in some
special situation that we now describe.

We can classify conformal field theory in two classes:

i) {\it open} conformal field theory, when the quantum conformal algebra
contains an infinite number of (generically non-conserved) currents;

ii) {\it closed} conformal field theory, when the quantum conformal algebra
closes with a finite set of (conserved) currents.

The FGGN theorem implies in particular that the spectrum is identically zero
if one $h_{2s}$ is zero, and identically infinity if one $h_{2s}$ is
infinity. Precisely:

\noindent a) if $h_{2s}=0$ for some $s>1$, then $h_{2s}=0$ $\forall s>0$, and

\noindent b)\ if $h_{2s}=\infty $ for some $s>1$, then $h_{2s}=\infty $ $%
\forall s>1$.

Equipped with this, we can describe the moduli space of conformal field
theory as a ball centred in free-field theory. As a radius $r$ one can take
the value of any $h_{2s}$ with $s>1$. The boundary sphere is the set of closed
theories. The bulk is the set of open theories.

Let us discuss the two cases $r=0$ and $r=\infty $ separately.

It is a rigorous and completely general consequence of the theorem
that when $r=\infty $ all current multiplets have infinite anomalous
dimensions and decouple from the OPE (with the only possible exception of $%
{\cal T}^{*}$, which is ``screened'' by ${\cal T}$). Supersymmetry plays an
important role here, since each multiplet necessarily has some component with
even spin, and therefore falls under the range of the Nachtmann theorem for $%
r\rightarrow \infty $.

The limit in which $r\rightarrow \infty $ is the limit of maximally strong
interaction, in the sense that once the quantum conformal algebra closes,
there is no way to make the interaction any stronger. It is not sufficient
to take $g\rightarrow \infty $:\ in N=4 supersymmetric Yang--Mills theory,
indeed, the $g\leftrightarrow 1/g$ duality suggests that the limit $%
g\rightarrow \infty $ at $N_{c}$ fixed is free and not closed. To have the
maximally strong interaction, one needs to take the large-$N_{c}$ limit at
the same time.

In the limit $r\rightarrow 0$ some currents with non-vanishing anomalous
dimension might survive, in principle, since $r$ is sensitive only to the 
{\it minimal} anomalous dimension of each even-spin level. It is
nevertheless reasonable to expect that $r\rightarrow 0$ reduces to a
free-field theory, and this is what we conjecture. Indeed, no interacting
theory with infinitely many conserved currents is known. An interesting
case, in this respect, is N=4 supersymmetric Yang--Mills theory, where the
spectrum $h_{2s}$ includes the full set of anomalous dimensions and
therefore $r\rightarrow 0$ ensures that all higher-spin currents generated
by the OPE are conserved.

The picture that has emerged can be summarized by the following statements.

$i)$ {\it Closed conformal field theory is the boundary of the moduli space
of open conformal field theory.}

$ii)$ {\it Closed conformal field theory is the exact solution to the
strongly coupled large-$N_{c}$ limit of open conformal field theory.}

$iii)$ {\it A closed quantum conformal algebra determines uniquely the
associated conformal field theory.}

$iv)$ {\it A closed quantum conformal algebra is determined uniquely by two
central charges, $c$ and $a$.}

We now comment on point $(iv)$. The basic procedure to determine the quantum
conformal algebra of closed conformal field theory (the so-called {\sl %
fusion rules}) can be applied to any set of finite operators (for example,
non-singlet currents with respect to some flavour group), although we focus
on the minimal algebra (namely the one of the stress tensor) for the sake of
generality. The procedure consists of the following steps. One studies the
free-field OPE of an open conformal field theory and organizes the currents
into orthogonal and mixing multiplets. Secondly, one turns a weak
interaction on and computes the anomalous dimensions of the operators to the
lowest orders in the perturbative expansion. Finally, one drops all the
currents with a non-vanishing anomalous dimension. More generically, one can
postulate a set of spin-0, 1 and 2 currents, which we call ${\cal T}_{0,1,2}$%
, and study the most general OPE algebra consistent with closure and unitary.

The closed N=2 quantum conformal algebra for generic $c$ and $a$ reads
schematically 
\begin{eqnarray*}
{\cal T}_{0}\,{\cal T}_{0} &=&{\frac{c}{|x|^{4}}}+{\frac{1}{|x|^{2}}}{\cal T}%
_{0}, \\
{\cal T}_{1}\,{\cal T}_{1} &=&{\frac{{c}}{|x|^{6}}}+\frac{1}{|x|^{4}}{\cal T}%
_{0}+\left( 1-{\frac{a}{c}}\right) \frac{1}{|x|^{3}}{\cal T}_{1}+\frac{1}{%
|x|^{2}}{\cal T}_{2}, \\
{\cal T}_{2}\,{\cal T}_{2} &=&{\frac{{c}}{|x|^{8}}}+\left( 1-{\frac{a}{c}}%
\right) {\frac{1}{|x|^{6}}}{\cal T}_{0}+\left( 1-{\frac{a}{c}}\right) {\frac{%
1}{|x|^{5}}}{\cal T}_{1}+\frac{1}{|x|^{4}}{\cal T}_{2},
\end{eqnarray*}
plus descendants and regular terms. We have emphasized those coefficients
that are proportional to $\left( 1-{a/c}\right) $. We observe that

$\circ $ {\it the $c=a$ closed algebra is unique and coincides with the N=4
one.}

$\circ$ {\it given $c$ and $a$, there is a unique closed conformal algebra
with N=2 supersymmetry.}

$c$ has a natural interpretation as the central extension of the algebra,
while the combination $(1-a/c)$ is a structure constant.

There might be a slightly more general, but still closed, structure, if the
multiplet ${\cal T}^{*}$, which mixes with ${\cal T}$, does not drop. This
algebra is determined by $c$, $a$ and the anomalous dimension of $%
{\cal T}^{*}.$

Finally, we observe that in N=1 (and non-supersymmetric) theories the
multiplet of the stress-tensor will not contain spin-0 partners, in general,
but at most the $R$-current. The above considerations stop at the spin-2 and
1 levels of the OPE, but the procedure to determine the closed algebra is
the same. What is more subtle is to identify the physical situation that the
closed limit should describe.

\vskip .3truecm \noindent \noindent {\bf 3. Quantum irreversibility} \vskip %
.2truecm

\noindent Quantum field theory defines a natural fibre bundle. The base
manifold is the space of physical correlators and the fibre is the space of
scheme choices, with suitable regularity restrictions. A projection onto the
base manifold is well defined and ensures scheme independence of the
physical correlators. We call this bundle the {\it scheme bundle}.

The scheme bundle is equipped with a metric $f$ and a fundamental one-form $%
\omega $, defined as follows. Consider the two-point function of the trace $%
\Theta $ of the stress tensor. In four dimensions, we normalize it as 
\[
\langle \Theta (x)~\Theta (0)\rangle =\frac{1}{15\pi ^{4}}\frac{\beta
^{2}(t)f(t)}{|x|^{8}}. 
\]
Reflection positivity ensures that $f\geq 0$. Actually, $f$ is strictly
positive throughout the RG flow, since the zeros of the two-point function
are parametrized precisely by the factor $\beta ^{2}$. Therefore 
$f$ is a
metric in the space of coupling constants, defined on the fibre. The beta
function is also defined on the fibre, since it is scheme-dependent, but the
combination $\beta ^{2}(t)f(t)$ is scheme-independent and therefore lives
on the base manifold. It is not a metric on the base manifold, however,
since it vanishes at the critical points.

The fundamendal one-form $\omega $ is defined as
\begin{equation}
\omega =-{\rm d}\lambda ~\beta (\lambda )~f(\lambda ),  \label{omega}
\end{equation}
$\lambda $ denoting the coupling constant, such that $\Theta =\beta (\lambda
){\cal O}$ for a suitable operator ${\cal O}$. 
In particular, $\lambda =\ln \alpha $
in a gauge field theory, where ${\cal O}=F^{2}/4$.

The central charge $a$ multiplies the Gauss--Bonnet integrand, or Euler
density,
\[
{\rm G}_{n}=(-1)^{\frac{n}{2}}\varepsilon _{\mu _{1}\nu _{1}\cdots \mu _{%
\frac{n}{2}}\nu _{\frac{n}{2}}}\varepsilon ^{\alpha _{1}\beta _{1}\cdots
\alpha _{\frac{n}{2}}\beta _{\frac{n}{2}}}\prod_{i=1}^{\frac{n}{2}}R_{\alpha
_{i}\beta _{i}}^{\mu _{i}\nu _{i}}, 
\]
in the trace anomaly coupled to an external gravitational field. ${\rm G}%
_{n} $ is a non-trivial total derivative, i.e. the total derivative of a
non-gauge-covariant current (the Chern--Simons form) and so it is defined up
to trivial total derivatives, the divergence of a gauge-covariant current.
The topological numbers calculated with a modified Gauss--Bonnet integrand
of the form $\tilde{{\rm G}}_{n}={\rm G}_{n}+\nabla _{\alpha }J^{\alpha }$
are exactly the same as those computed with ${\rm G}_{n}$.

The modified integrand can be chosen to be linear in the conformal
factor, and in that case we call it {\it pondered} Euler density. In
particular $\tilde{{\rm G}}_{n}\propto \Box ^{\frac{n}{2}}\phi $ for a
conformally-flat metric $g_{\mu \nu }=\delta _{\mu \nu }{\rm e}^{2\phi }$.
Writing $\sqrt{g}{\rm G}_{n}=\partial _{\alpha }C^{\alpha },$ where $%
C^{\alpha }$ is the Chern--Simons form, the {\it pondered} Chern--Simons
form reads $\tilde{C}^{\alpha }=C^{\alpha }+\sqrt{g}J^{\alpha }$ and $\sqrt{g%
}{\rm \tilde{G}}_{n}=\partial _{\alpha }\tilde{C}^{\alpha }.$

In four dimensions, $\tilde{{\rm G}}_{4}$ reads 
$
\tilde{{\rm G}}_{4}={\rm G}_{4}-{\frac{8}{3}}\Box R={\rm G}_{4}+\nabla
_{\alpha }J_{4}^{\alpha }$, with 
$J_{4}^{\alpha }=-{\frac{8}{3}}\nabla_{\alpha }R$.
In six dimensions we have $\tilde{{\rm G}}_{6}\equiv {\rm G}_{6}+\nabla
_{\alpha }J_{6}^{\alpha }$ with 
\[
J_{6}^{\alpha }=-\frac{48}{5}R^{\alpha \mu }\nabla _{\mu }R+{\frac{102}{25}}%
\nabla ^{\alpha }R^{2}-12\nabla ^{\alpha }(R_{\mu \nu }R^{\mu \nu })-\frac{24%
}{5}\nabla ^{\alpha }\Box R,
\]
so that on conformally-flat metrics 
$\sqrt{g}\,\tilde{{\rm G}}_{4}=16\Box^2\phi$ and
$\sqrt{g}\,\tilde{{\rm G}}_{6}=48~\Box ^{3}\phi$.

In generic $n$ the pondered Euler density has the form 
\begin{equation}
\tilde{{\rm G}}_{n}={\rm G}_{n}+\nabla _{\alpha }J_{n}^{\alpha }={\rm G}%
_{n}+\cdots +p_{n}\Box ^{n/2-1}R,\qquad J_{n}^{\alpha }=\cdots +p_{n}\nabla
^{\alpha }\Box ^{{\frac{n}{2}}-2}R,  \label{tyco}
\end{equation}
and on conformally-flat metrics $\sqrt{g}\tilde{{\rm G}}_{n}=-2(n-1)p_{n}%
\Box ^{n}\phi $. Only the coefficient $p_{n}$ in (\ref{tyco}) is relevant
for us and the definition of $\tilde{{\rm G}}_{n}$ makes it easily
calculable.

The Euler characteristic of the $n$-dimensional sphere $S^{n}$ is equal to
2. In our notation we can write 
\[
(-1)^{\frac{n}{2}}2^{\frac{3n}{2}+1}\pi ^{\frac{n}{2}}\left( {\frac{n}{2}}%
\right) !=\int_{S^{n}}\sqrt{g}{\rm G}_{n}\,{\rm d}^{n}x=\int_{S^{n}}\sqrt{g}%
\tilde{{\rm G}}_{n}\,{\rm d}^{n}x=-2(n-1)p_{n}\int_{S^{n}}\Box ^{\frac{n}{2}%
}\phi \,{\rm d}^{n}x.
\]
The calculation in the sphere with 
metric ${\rm d}s^{2}={\frac{{\rm d}x^{2}}{(1+x^{2})^{2}}}$,
gives $p_{n}=-{\frac{2^{\frac{n}{2}}n}{2(n-1)}}$,
which agrees with the known values in $n=4$ and $n=6$. In \cite{6d} the
expression of $\tilde{{\rm G}}_{8}$ is also worked out and $p_{8}$ is
checked.

Summarizing, on conformally flat metrics 
\[
\sqrt{g}\tilde{{\rm G}}_{n}=2^{\frac{n}{2}}n\,\Box ^{\frac{n}{2}}\phi . 
\]

According to ref. \cite{athm} the Euler density that should appear in the
trace anomaly should be precisely the pondered Euler density, thereby
removing the ambiguities associated with the coefficients $a'$ of the
trivial total derivative terms
of the form $\nabla _{\alpha }J_{n}^{\alpha }$. The dependence of the trace
anomaly on the conformal factor $\phi $ becomes extraordinarily simple:
\begin{equation}
\Theta =a_{n}\,\tilde{{\rm G}}_{n}+{\rm conf.\,invs.}=2^{\frac{n}{2}%
}n\,a_{n}\,{\rm e}^{-n\phi }\Box ^{\frac{n}{2}}\phi,  \label{uno}
\end{equation}
and the relation between the total $a$-flow and the $\Theta $ two-point
function becomes manifest. Normalizing $a$ as in (\ref{uno}), the two-point
function reads at criticality 
\[
\langle \Theta (x)\,\Theta (y)\rangle =-2^{{\frac{n}{2}}}n\,a_{n}\,\Box ^{%
\frac{n}{2}}\delta (x-y) 
\]
and the expression for the $a$-flow is therefore:
\begin{equation}
a_{n}^{{\rm UV}}-a_{n}^{{\rm IR}}={\frac{\int {\rm d}^{n}x\,|x|^{n}\,\langle
\Theta (x)\,\Theta (0)\rangle }{2^{{\frac{3n}{2}}-1}\,n\,\Gamma (n+1)}}.
\label{6d}
\end{equation}

A convenient normalization of $a_{n}$ is that it be equal to 
1 for a real scalar
field, and reads in general 
\begin{equation}
N_{s}+f_{n}N_{f}+v_{n}N_{v}  \label{norma}
\end{equation}
for free-field theories with $N_{s}$ real scalar fields, $N_{f}$ Dirac
fermions and $N_{v}$ vectors. In $n=4$ we change the normalization of $a$
according to this convention ($f_{4}=11,$ $v_{4}=62$) and write 
\begin{equation}
\Delta a=a_{{\rm UV}}^{{}}-a_{{\rm IR}}^{{}}=\int_{{\rm UV}}^{{\rm IR}%
}\omega \geq 0.  \label{formula}
\end{equation}
The total RG\ flow of $a$ is the integral of the fundamental one-form $%
\omega $ between the fixed points. Quantum irreversibility is measured by
the invariant (i.e. scheme-independent) area of the graph of the beta
function between the fixed points. In this integral, scheme independence is
re\-pa\-ra\-me\-tri\-za\-tion invariance.

This formula can be checked to the fourth-loop order in the most general
renormalizable theory. Here we focus on QCD in the conformal window in the
neighbourhood of the asymptotic freedom point $N_{f}={\frac{11}{2}}N_{c}$.
The strategy for computing higher-loop corrections to the trace anomaly was
developed in refs. \cite{brown}. See \cite{6d} for its extension to six
dimensions.

Collecting the results of these references in a general formula, the
third-loop RG flow of $a$ reads 
\[
a_{{\rm UV}}-a_{{\rm IR}}={\frac{1}{2}}f_{{\rm UV}}\beta _{2}\alpha _{{\rm IR%
}}^{2}+{\cal O}(\alpha _{{\rm IR}}^{3}),
\]
where $\beta _{1}$ and $\beta _{2}$ are the first two coefficients of the
beta function, $\beta (\alpha )=\beta _{1}\alpha +\beta _{2}\alpha ^{2}+%
{\cal O}(\alpha ^{3})$. Formula (\ref{formula}) gives exactly the same
result. Concretely, with $N_{f}$ flavours and $N_{c}$ colours we have 
\[
a_{{\rm UV}}-a_{{\rm IR}}={\frac{44}{5}}N_{c}N_{f}\left( 1-{\frac{11}{2}}{%
\frac{N_{c}}{N_{f}}}\right) ^{2}.
\]

In supersymmetric QCD the prediction can be compared with the exact formula (%
\ref{fora}). The check can be extended to the fourth-loop order \cite{athm},
in both the supersymmetric and non-supersymmetric cases. \noindent In six
dimensions, formula (\ref{6d}) has been checked to the fourth-loop order in
the theory $\varphi ^{3}$ \cite{6d}.

\medskip

I now discuss the extension of these results to all orders. 
Renormalization can be seen \cite{athm} as the restoration of positivity 
(or, better, boundedness from below) 
of the generating functional of 1PI
diagrams in the Euclidean framework. This positivity
is in general violated by the regularization procedure and divergences.

Here, we consider the induced action for the conformal factor $\phi$. 
Despite the fact that $\phi$ is an external source, 
the positivity property holds because $\phi$ couples to $\Theta$, an
evanescent operator. At most, we might have to adjust the unique
free parameter (``coupling constant'') at our disposal: the $a'$ ambiguity.

The quantum-irreversibility formula is derivable from the statement:

\noindent {\it the induced effective action $S_{{\rm R}}$ for the conformal
factor $\phi $ \cite{riegert} is positive-definite throughout the RG flow,
if and only if it is positive-definite at a given energy,}

\noindent which implies the ``$a$-theorem'':

i) $a$ {\it is non-negative;}

ii) {\it the total RG flow of $a$ is non-negative and equal to the invariant
area of the beta function:} 
\[
a_{{\rm UV}}-a_{{\rm IR}}=-\int_{\lambda _{{\rm UV}}}^{\lambda _{{\rm IR}}}%
{\rm d}\lambda \ \beta (\lambda )f(\lambda )\geq 0.
\]

Now, $S_{{\rm R}}[\phi ]$ is the solution of the equation 
$
\Theta ={\rm e}^{-4\phi }\frac{\delta S_{{\rm R}}[\phi ]}{\delta \phi }$.
At criticality in the Euclidean framework we have 
\[
\Theta =\frac{1}{90(4\pi )^{2}}\left[ a_{*}{\rm e}^{-4\phi }\Box ^{2}\phi +%
\frac{1}{6}(a_{*}-a_{*}^{\prime })\Box R\right] 
\]
and therefore 
\[
S_{{\rm R}}[\phi ]=\frac{1}{180}\frac{1}{(4\pi )^{2}}\int {\rm d}%
^{4}x\,\{a_{*}(\Box \phi )^{2}-(a_{*}-a_{*}^{\prime })\left[ \Box \phi
+(\partial _{\mu }\phi )^{2}\right] ^{2}\}.
\]
The two terms of $S_{{\rm R}}[\phi ]$ have to be separately positive. In
particular, positivity of the first term implies $a_{*}>0$ in the IR if $%
a_{*}>0$ in the UV. This is true, since $a_{{\rm free}}>0$ in a free-field
theory.

The quantity $a^{\prime }$ is defined up to an additive, 
coupling-independent
constant and needs to be normalized at a given energy scale. 
The quantity whose RG flow is given by (\ref
{formula}) is precisely $a^{\prime }$ and our statement amounts 
to showing that $\Delta a^{\prime }$,
which is certainly non-negative, is equal to $\Delta a$.

The second term of $S_{{\rm R}}[\phi ]$ is positive at criticality if 
$
a_{*}^{\prime }\geq a_{*}$.  
This condition has to hold throughout the renormalization group flow, in
particular 
$
a^{\prime}_{{\rm UV}}\geq a_{{\rm UV}}
$ if and only if $a^{\prime}_{{\rm IR}}\geq a_{{\rm IR}}$. 
Now, we know that $a_{{\rm UV}}^{\prime }\geq a_{{\rm IR}}^{\prime }$. Let
us fix $a^{\prime }$ by demanding that $a$ and $a^{\prime }$ coincide in the
UV, $a_{{\rm UV}}^{\prime }=a_{{\rm UV}}$. Then we have, combining the
various inequalities derived so far,
$
a_{{\rm UV}}=a_{{\rm UV}}^{\prime }\geq a_{{\rm IR}}^{\prime }\geq a_{{\rm IR%
}}$
from which the claimed inequality $a_{{\rm UV}}\geq a_{{\rm IR}}$ follows.

Now, let us tentatively suppose that with the normalization $a_{{\rm UV}%
}^{\prime }=a_{{\rm UV}}$ we have the strict inequality $a_{{\rm IR}%
}^{\prime }>a_{{\rm IR}}$. We prove that this is absurd and conclude that $%
a_{{\rm IR}}^{\prime }=a_{{\rm IR}}$.

We can do this by changing the normalization of $a^{\prime }$ with the shift 
$a^{\prime }\rightarrow a^{\prime ~{\rm new}}=a^{\prime }-a_{{\rm IR}%
}^{\prime }+a_{{\rm IR}},$ so that $a_{{\rm IR}}^{\prime ~{\rm new}}=a_{{\rm %
IR}}$. We have $a_{{\rm UV}}^{\prime }\rightarrow $ $a_{{\rm UV}}^{\prime ~%
{\rm new}}=a_{{\rm UV}}^{\prime }-a_{{\rm IR}}^{\prime }+a_{{\rm IR}}$ and
therefore $a_{{\rm UV}}^{\prime }$ no longer satisfies the inequality
$a_{*}^{\prime }\geq a_{*}$, since $a_{{\rm UV}}^{\prime ~{\rm new}}<a_{{\rm UV}}$. This is a
contradiction. We conclude that 
$
a_{{\rm UV}}^{\prime }=a_{{\rm UV}}$ if and only if $a_{{\rm IR}%
}^{\prime }=a_{{\rm IR}}$.

\medskip

These arguments are somewhat orthogonal, or complementary, to the approach 
{\sl \`{a} la} spectral representation of \cite{cappelli}. In particular,
knowledge about the (positivity) properties of the local parts of the
correlators is of fundamental importance.

\medskip

An extension of these ideas to odd dimensions, which is not straightforward
since there is no trace anomaly in external gravity in odd dimensions, can
be obtained by dimensional continuation. The resulting formula is testable,
in principle, in models interpolating between pairs of free-field fixed
points.

\medskip

Finally, I stress that the phenomenon of quantum irreversibility is proper
to the dynamical scale $\mu $, i.e. it is the intrinsic drift
of the renormalization group. Explicit scales (masses, Newton's constant and
other dimensionful parameters) need not be described by formula (\ref
{formula}). Moreover, formula (\ref{formula}) has to be replaced by a more
complicated expression also when the stress tensor is not truly finite, but
mixes with other operators (a well-known example
is the $\lambda \varphi ^{4}$-theory
- see sect. 2.3 of \cite{athm} for details). These are signals of the richness
of higher-dimensional conformal field theories\ with respect to the two-dimensional
ones. There is, nevertheless, a remarkable subset, the set $c=a$, where the
two-dimensional properties are best reproduced. This set admits a
generalization to arbitrary dimension and it is defined as the set of
conformal field theories having a trace anomaly quadratic in the Ricci
tensor and Ricci curvature \cite{c=a}. 

\medskip

I thank J. Erlich, D.Z. Freedman, M. Grisaru 
and A.A. Johansen for collaboration of the first topic of this research.

\end{document}